\documentclass[12pt]{article}
\usepackage[]{epsfig}
\usepackage[centertags]{amsmath}
\usepackage{amsfonts}
\usepackage{amssymb}
\usepackage[english]{babel}
\usepackage{amsmath, amsthm, slashed,url}
\usepackage{color}
\begin{document}

\title{\bf Magnetic superconfinement of  Dirac fermions zero-energy modes in bilayer graphene quantum dots}
\author{ 
Lucas~Sourrouille
\\
{\normalsize \it IFISUR, Departamento de F\'isica (UNS-CONICET) }\\
{\normalsize\it  Avenida Alem 1253, Bah\'ia Blanca, 8000, Buenos Aires, Argentina }
\\
{\footnotesize  sourrou@df.uba.ar} } \maketitle

\abstract{We show that in bilayer graphene it is possible to achieve a very restrictive confinement of the massless Dirac fermions zero-modes by using inhomogeneous
magnetic fields. Specifically, we show that, using a suitable nonuniform magnetic fields, the wave function may be restricted to a specific region of the space, being forbidden all transmission probability to the contiguous regions. This allows to construct mesoscopic structures in bilayer graphene by  magnetic fields configurations.

}
 
\vspace{0.3cm}


\vspace{1cm}
\section{Introduction}
The experimental realization of monolayer graphene films \cite{1.1,2.1,3.1} has allowed explore the physics of two-dimensional (2D) Dirac-Weyl fermions. The study of the influence of magnetic fields on such quasi-particles has a great interest both from fundamental and applied points of views.
Let us just mention, for example, the Landau levels and Hall
effect and the design of nanoelectronic divices \cite{martino07,peeters09,portnoi16a,portnoi16b}. Nevertheless, in contrast to Schr\"{o}dinger case, Dirac fermions
can penetrate electrostatic barriers with high
transmission probability, leading to a great difficulty in
confining electrons electrostatically \cite{8,9,10}.  This difficulty has given rise to consider pure
magnetic traps \cite{11,12,13,14}. Many kinds of magnetic confinement have been treated, for instance with inhomogeneous field profiles \cite{22,23,24,kuru09,26,27,28}, magnetic antidots \cite{martino07} or anti-rings \cite{31,34}. In particular, magnetic traps composed by fields with a slowly decaying nature, was studied in \cite{portnoi16a,portnoi16b,14}. However, in this kind of magnetic confinement, the wave function is not restricted to a specific region of the space, e.g., a finte region, not being forbidden  transmission probablity to all the space. 
\\
Here, we propose a new confining mechanism in graphene. Firstly, we analyze the solutions of the Dirac-Wely equation in the presences of magnetic fields whose behavior is dictated by $B(r) = b r^{\alpha}$, with $b$ and $\alpha$ are arbitrary real numbers. We will pay attention to the zero energy levels. We show that if $\alpha > -2$ and $b>0$ the spectrum of the angular momentum for the zero modes satisfies $\ell \leq 0$, whereas if $b<0$ we have $\ell \geq -1$. Here, we denote with $\ell$ the eigenvalues of the angular momentum. The case $\alpha = -2$ is particular and the spectrum of the angular momentum is extended to all integers, regardless of the sign of $b$. Finally, we find that if $\alpha < -2$ the angular momentun is restricted to  $\ell 
\geq 1$ and $\ell \leq -2$ for $b>0$ and $b<0$ respectively.
We use these ideas to study massless Dirac fermions zero-modes in the presences of a magnetic quantum dot, defined by, 
\begin{equation}
B(r) = \left\{\begin{array}{ll}
b_1,\qquad &r<r_0
\\[1.5ex]
b_2 r^{-3},\qquad &r>r_0\,,
\end{array}\right.\,.
\label{}
\end{equation}
being $b_1$ and $b_2$ are real constants, related by $b_1 = b_2 r_0^{-3}$. Then, by using the previous ideas, we show that  there are no solutions of zero energy for single layer graphene. We  
also analyze the case of bilayer graphene. By doing a similar analysis we conclude that there is no possible to find zero energy modes that coexist in both regions of the space. However, this does not imply that there are no solutions of zero energy. As we will see, we can construct eigenstates which are different from zero only in one of the two regions of the space. We will call this kind of confinement as superconfinement because it is  more restrictive than the well know magnetic confinement in graphene \cite{martino07}-\cite{portnoi16b},\cite{22}-\cite{34}.
\section{Single-layer graphene }
Let us start by considering
the quasi-particles Hamiltonian describing the 2D excitations in graphene 
\begin{equation}
H= \upsilon_F \sigma^i p_i = \upsilon_F (\sigma^x p_x +  \sigma^y p_y)\;,
\label{}
\end{equation}
Here, the $\sigma^i$ 
are 2$\times$2 Pauli matrices, i.e.
\begin{eqnarray}
\sigma^1 =\left( \begin{array}{cc}
0 & 1 \\
1 & 0 \end{array} \right)
\,,
\;\;\;\;\;\
\sigma^2 =\left( \begin{array}{cc}
0 & -i \\
i & 0 \end{array} \right)\;,
\end{eqnarray}
$\upsilon_F$ is the Fermi velocity and $p_i =-i\partial_i$ is the two-dimensional momentum operator. 
In a perpendicular
magnetic field, $B = \nabla \times A$, we can represent the effect of the field by
the potential vector ${\bf A}$, leading us to the Dirac equation 
\begin{equation}
H \Psi (x, y) = \upsilon_F\sigma^i D_i \Psi (x, y) = E \Psi (x, y)
\label{2dw}
\end{equation}
where, $D_{i}= -i\partial_{i} +A_{i}$ $(i =1,2)$ is the the covariant derivative and $\Psi (x, y, t)$ is the two-component spinor
\begin{equation}
\Psi=(\psi_a,\psi_b)^T
\label{}
\end{equation}
The general form of the vector potential is 
$
{\bf A} = \frac{f(r)}{2}\,(-{y}/{r} , {x}/{r} ) 
$, 
 which can be rewritten in polar coordinates as 
\begin{equation}\label{a}
{\bf A} =  \frac{f(r)}{2} \,\hat \theta,\qquad \frac{f(r)}{2} \equiv A_\theta(r) \,,
\end{equation}
where $\hat \theta$ is the unit vector in the azimuthal direction.
It is evident that this vector potential is in the Coulomb gauge, i.e.
%
$
\nabla\cdot {\bf A}= \frac{1}{r}\partial_\theta f(r) =0
$, 
thus it is possible to introduce an scalar field $\lambda(r)$ such that
\begin{equation}
\frac{f(r)}{2} = \partial_r \lambda(r),\qquad \,.
\end{equation}
In our case, due to the axial symmetry,  we write this Hamiltonian
in polar coordinates $(x,y) = (r\cos \theta, r \sin \theta)$, so that
the expression of $H$ in (\ref{2dw}) becomes
\begin{equation}
H = \upsilon_F
\left(
\begin{array}{cc}
 0 & e^{-i\theta}\left( -i\partial_r +\frac{\partial_\theta}{r} + \frac{i}{2} f(r)\right)
 \\[2.ex]
e^{i\theta}\left(-i\partial_r -\frac{\partial_\theta}{r} - \frac{i}{2} f(r)\right) & 0
\end{array}
\right) \label{eq0}
\end{equation}
where we have taken into account the notation (\ref{a}) for the magnetic vector potential.
This Hamiltonian commutes with the operator $J_z =
L_z + \sigma_z/2$, where $L_z = x\partial_y - y\partial_x= -i\partial_\theta$ is the z-component of the
orbital angular momentum. As a consequence, we can choose the eigenstates in the form
\begin{equation}
\Psi(r,\theta) = \left(\begin{array}{c}
\psi_1(r,\theta)\\[1ex]
\psi_2(r,\theta)\end{array}\right) 
=\left(\begin{array}{c}
\xi(r) e^{i\ell\theta}\\[1ex]
i \chi(r) e^{i(\ell+1)\theta}\end{array}\right)
\label{eq1}
\end{equation}
where $\ell$ is for the integer eigenvalues of $L_z$, and the subindex 1, 2 of the spinor components
refer to the sublattices A, B, respectively, of the graphene honeycomb lattice.
Since the wave function (\ref{eq1}) is an
eigenfunction of the total angular momentum operator we have,
\begin{equation}
J_z\Psi = (L_z + \frac12 \sigma_z)\Psi = j \Psi\,.
\end{equation}
with the eigenvalue $j=\ell+1/2$.
Then Eq.(\ref{2dw}) reduces to a pair of coupled 1D equations for the radial part of the spinor components $\xi(r)$ and $\chi(r)$,
\begin{eqnarray}
\Big[\partial_r -\frac{\ell+1}{r} - \frac{f(r)}{2} \Big] \chi(r) = \varepsilon \xi(r)
\nonumber \\[3mm]
-\Big[\partial_r +\frac{\ell}{r} + \frac{f(r)}{2} \Big] \xi(r) = \varepsilon \chi(r)
\label{}
\end{eqnarray}
where the eigenvalue $\varepsilon = \frac{E}{\upsilon_F}$. 
\\
Let us now concentrate on solving these equations for different magnetic fields configurations. In particular, we concern for 
the solutions at zero energy,
\begin{eqnarray}
\Big[\partial_r -\frac{\ell+1}{r} - \frac{f(r)}{2} \Big] \chi(r) = 0
\nonumber \\[3mm]
\Big[\partial_r +\frac{\ell}{r} + \frac{f(r)}{2} \Big] \xi(r) = 0
\label{eq11}
\end{eqnarray}
One simple case consist on constant magnetic field which leads to the following field 
equations,
\begin{eqnarray}
\Big[\partial_r -\frac{\ell+1}{r} - \frac{b r}{2} \Big] \chi(r) = 0
\nonumber \\[3mm]
\Big[\partial_r +\frac{\ell}{r} + \frac{b r}{2} \Big] \xi(r) = 0
\label{eq12}
\end{eqnarray}
where, $b$ is a real number, equal to the magnetic field, $B =b$. The solutions are
\begin{eqnarray}
\chi(r) = r^{\ell +1} e^{br^2/4}
\nonumber \\[3mm]
\xi(r) = r^{-\ell} e^{-br^2/4}
\label{eq13}
\end{eqnarray}
Since the eigenstates (\ref{eq1}) must be regular at the origin, we require that 
\begin{eqnarray}
\lim_{r \to 0} \Psi(r,\theta) =0
\label{eq14}
\end{eqnarray}
In addition the eigenstates (\ref{eq1}) must be  normalizable, which implies that
\begin{eqnarray}
\lim_{r \to \infty} \Psi(r,\theta) =0
\label{eq15}
\end{eqnarray}
These conditions imply that zero-energy solutions can exist only for one (pseudo)spin direction, depending on the sign of the 
constant $b$. Thus, if $b>0$, we have,
\begin{equation}
\Psi(r,\theta)  
=\left(\begin{array}{c}
e^{i\ell\theta} r^{-\ell} e^{-br^2/4} \\[1ex]
0 \end{array}\right) \,,
\;\;\;\;\;\ \ell \leq 0
\label{eq16}
\end{equation}
On the other hand, the condition $b<0$ implies 
\begin{equation}
\Psi(r,\theta)
=\left(\begin{array}{c}
0\\[1ex]
i e^{i(\ell+1)\theta}r^{\ell +1} e^{br^2/4} \end{array}\right) \,,
\;\;\;\;\;\ \ell \geq -1
\label{eq17}
\end{equation}
If a magnetic field is chosen to be $B(r)=br^{-1}$, then it is not difficult to show that the solutions of the equation 
(\ref{eq11}) are,
\begin{equation}
\Psi(r,\theta)  
=\left(\begin{array}{c}
e^{i\ell\theta} r^{-\ell} e^{-br/2} \\[1ex]
0 \end{array}\right) \,,
\;\;\;\;\;\ \ell \leq 0 \;,
\label{}
\end{equation}
if $b>0$, and 
\begin{equation}
\Psi(r,\theta)
=\left(\begin{array}{c}
0\\[1ex]
i e^{i(\ell+1)\theta}r^{\ell +1} e^{br/2} \end{array}\right) \,,
\;\;\;\;\;\ \ell \geq -1 \;,
\label{}
\end{equation}
if $b<0$.
\\
In general, we can show that for magnetic fields of the form $B(r)=br^{\alpha}$, with $\alpha > -2$, the solutions exist only 
for one (pseudo)spin direction, depending on the sign of the constant $b$. In addition, it is interesting to note that the 
restriction on the values of the angular momentum is maintained for any $\alpha$ satisfying the condition $\alpha > -2$, i.e, 
if $b>0$ the spectrum of the angular momentum satisfies $\ell \leq 0$, whereas if $b<0$ we have $\ell \geq -1$. Notably, the 
situation is different for  $\alpha = -2$ and  $\alpha < -2$. By solving the equations (\ref{eq11}), for the case $\alpha = -2$, 
we have, 
\begin{eqnarray}
\chi(r) = r^{\ell +1} e^{b(\ln r)^2/2}
\nonumber \\[3mm]
\xi(r) = r^{-\ell} e^{-b(\ln r)^2/2}
\label{eq20}
\end{eqnarray}
In order to study the behavior of these functions as $r\to 0$ and $r\to \infty$ we can rename $\ln r$ as $z$. Then, the functions 
(\ref{eq20}) take a more simple form,
\begin{eqnarray}
\chi(r) = e^{z(\ell +1)} e^{b(z)^2/2}
\nonumber \\[3mm]
\xi(r) = e^{-z\ell} e^{-b(z)^2/2}
\label{eq21}
\end{eqnarray}
Assuming $b>0$, it is not difficult to check from (\ref{eq21}) the following boundary conditions,
\begin{eqnarray}
\lim_{r \to 0} \xi(r) = 0 \,,
\;\;\;\;\;\ \lim_{r \to 0} \chi(r) = \infty
\label{eq22}
\end{eqnarray}
\begin{eqnarray}
\lim_{r \to \infty} \xi(r) = 0 \,,
\;\;\;\;\;\ \lim_{r \to \infty} \chi(r) = \infty
\label{eq23}
\end{eqnarray}
Therefore, the only solution physically acceptable is $\xi(r) = e^{-z\ell} e^{-b(z)^2/2}$. Since, the conditions 
(\ref{eq22}) and (\ref{eq23}) are satisfied independently of the values of $\ell$, the spectrum of angular momentun is all integers.
Consequently, the eigenstate (\ref{eq1}) becomes, 
\begin{equation}
\Psi(r,\theta)  
=\left(\begin{array}{c}
e^{-z\ell} e^{-b(z)^2/2} e^{i\ell\theta}\\[1ex]
0 \end{array}\right) \,,
\;\;\;\;\;\ -\infty \leq \ell \leq +\infty \;,
\label{}
\end{equation}
Similar considerations may be done for the case $b<0$, in which case we obtain,
\begin{equation}
\Psi(r,\theta)
=\left(\begin{array}{c}
0\\[1ex]
i e^{i(\ell+1)\theta} e^{z(\ell +1)} e^{b(z)^2/2} \end{array}\right) \,,
\;\;\;\;\;\ -\infty \leq \ell \leq +\infty \;,
\label{}
\end{equation}
In order to ilustrate the solutions for situation $\alpha < -2$, we can consider $\alpha = -3$, then the solutions of the field equations (\ref{eq11}) are
\begin{eqnarray}
\chi(r) = r^{\ell +1} e^{b r^{-1} }
\nonumber \\[3mm]
\xi(r) = r^{-\ell} e^{-b r^{-1} }
\label{eq26}
\end{eqnarray}
Assuming $b>0$ and $\ell \geq 0$, the solutions (\ref{eq26}) have the following boundary conditions, 
\begin{eqnarray}
\lim_{r \to 0} \xi(r) = 0 \,,
\;\;\;\;\;\ \lim_{r \to 0} \chi(r) = \infty
\label{eq28}
\end{eqnarray}
\begin{eqnarray}
\lim_{r \to \infty} \xi(r) = 0 \,,
\;\;\;\;\;\ \lim_{r \to \infty} \chi(r) = \infty
\label{eq29}
\end{eqnarray}
Thus, the eigenstates of the Hamiltonian (\ref{eq0}) must be, 
\begin{equation}
\Psi(r,\theta)  
=\left(\begin{array}{c}
r^{-\ell} e^{-b r^{-1}} e^{i\ell\theta}\\[1ex]
0 \end{array}\right) \,,
\;\;\;\;\;\ \ell \geq 1 \;,
\label{eq29}
\end{equation}
Here, it is important to note that for the existence of solutions, it is necessary to impose the condition 
$\ell \geq 1$, otherwise, the solution would not be normalized and therefore there would not be solutions other than the 
trivial one.
\\
On the other hand, if $b<0$, the eigenstates of the Hamiltonian are 
\begin{equation}
\Psi(r,\theta)  
=\left(\begin{array}{c}
0\\[1ex]
i e^{i(\ell+1)\theta} r^{\ell +1} e^{b r^{-1} } \end{array}\right) \,,
\;\;\;\;\;\ \ell \leq -2 \;,
\label{eq30}
\end{equation}
It is not difficult to imagine a generalization of this result for other values satisfying $\alpha < -2$,
\begin{equation}
\Psi(r,\theta)  
=\left(\begin{array}{c}
r^{-\ell} e^{-\frac{b}{(\alpha +1)^2} r^{\alpha+1}} e^{i\ell\theta}\\[1ex]
0 \end{array}\right) \,,
\;\ \ell \geq 1 \,,
\;\ b>0 \;,
\label{}
\end{equation}
\begin{equation}
\Psi(r,\theta)  
=\left(\begin{array}{c}
0\\[1ex]
i e^{i(\ell+1)\theta} r^{\ell +1} e^{\frac{b}{(\alpha +1)^2} r^{\alpha+1} } \end{array}\right) \,,
\;\ \ell \leq -2 \,,
\;\ b<0 \;,
\label{}
\end{equation}
Hence, we can classify the spectrum of angular momentums into three different classes. For instance, if $b>0$, we have for 
$\alpha > -2$ an spectrum of angular momentums that satisfy the condition $\ell \leq 0$, if $\alpha = -2$ all 
integers are allowed for the spectrum, while, if $\alpha < -2$, the possible values of angular momentums should satisfy $\ell 
\geq 1$. If $b<0$, we must replace $\ell \leq 0$ by $\ell \geq -1$ and $\ell \geq 1$ by $\ell \leq -2$. From this, we can 
conclude that the solutions of the Dirac equation (\ref{2dw}) for an electron in a magnetic field $B(r)=br^{\alpha}$ with $\alpha 
> -2$ do not have common values in the spectrum of angular momentum with the solutions for an electron in a magnetic field 
with $\alpha < -2$.
\\
By using these ideas we will construct a magnetic field configuration and we will show that there are no solutions of zero energy for single layer graphene. In addition, we will conclude that there is no possible to find zero energy modes that coexist in both regions of the space. However, this does not imply that there are no solutions of zero energy. As we will see, we can construct eigenstates which are different from zero only in one of the two regions of the space.
\\
Let us consider a magnetic field configuration, defined by 
\begin{equation}
B(r) = \left\{\begin{array}{ll}
b_1,\qquad &r<r_0
\\[1.5ex]
b_2 r^{-3},\qquad &r>r_0\,,
\end{array}\right.\,.
\label{eq33}
\end{equation}
Here, $b_1$ and $b_2= b_1 r_0^{-3}$ are real constants.
Let us  analyze the possible solutions of Eq.(\ref{eq11}) in region I, i.e. $r<r_0$. In virtue of the solutions (\ref{eq16}) and (\ref{eq17}), there are only two possible solutions of zero energy.  
The wave
functions in region II, i.e. $r>r_0$, can be found in the same way. We can conclude that the only possible eigenstates are the spinors (\ref{eq29}) and (\ref{eq30}).
If an eigenstate of the Dirac equation exists in two regions, both wave function components should be continuous across the interface at $r = r_0$, that is, we should be able to match both functions (\ref{eq16}) with (\ref{eq29}) in the case of $b>0$ and (\ref{eq17}) with (\ref{eq30}) for $b<0$ . However, this is not possible, since the  angular momentum of the eigenstate (\ref{eq16}) and (\ref{eq17}) are subject to the restrictions $\ell \leq 0$ and $\ell \geq -1$ respectively, whereas, the angular momentum of the spinor (\ref{eq29}) and (\ref{eq30}) satisfy $\ell \geq 1$ and $\ell \leq -2$ respectively. This suggests that the  wave function can not coexist in both regions of the space. Another possibility would be consider a superconfinement of the wave function, that is, we could assume that the wave function is different from zero only in one of the two regions. This requires that the wave function vanishes at $r=r_0$. However, as we see from solutions (\ref{eq16}) and (\ref{eq17}), and (\ref{eq29}) and (\ref{eq30}) this is not possible. Thus, there are no solutions of zero energy, besides the trivial one. 
\section{Magnetic superconfinement in bilayer graphene}
Let us concentrate on bilayer graphene. The bilayer 
graphene \cite{mc}-\cite{mm} in the simplest approximation can
be considered as a zero-gap semiconductor with parabolic
touching of the electron and hole bands described by the
single-particle Hamiltonian.
By exfoliation of graphene one can obtain several layers of carbon atoms.
In this simple approximation the Hamiltonian of the bilayer graphene is,
\begin{equation}
H = \upsilon_F\left( \begin{array}{cc}
0 & (D_1 -iD_2)^2 \\
(D_1 +iD_2)^2 & 0 \end{array} \right)
\label{3dw3}
\end{equation}
In polar coordinates the Hamiltonian is written as
\begin{equation}
H = \upsilon_F
\left(
\begin{array}{cc}
 0 & e^{-i2\theta}\left( -i\partial_r +\frac{\partial_\theta}{r} + \frac{i}{2} f(r)\right)^2
 \\[2.ex]
e^{i2\theta}\left(-i\partial_r -\frac{\partial_\theta}{r} - \frac{i}{2} f(r)\right)^2 & 0
\end{array}
\right) \label{eq36}
\end{equation}
Hence, the field equations for the zero energy states in a constant magnetic field are, 
\begin{eqnarray}
\Big[-i\partial_r + \frac{\partial_\theta}{r} + i\frac{b r}{2} \Big]^2 \chi(r) e^{i(\ell +1)\theta}= 0
\nonumber \\[3mm]
\Big[-i\partial_r -\frac{\partial_\theta}{r} -i \frac{b r}{2} \Big]^2 \xi(r)e^{i\ell\theta} = 0
\label{eq37}
\end{eqnarray}
One can see immediately from Eq. (\ref{eq37}) that there are zero modes
and their number is twice as great as for the case of a single layer \cite{katnelson}-\cite{6k}.
Indeed, for $B= b>0$, 
\begin{equation}
\Psi(r,\theta)  
=\left(\begin{array}{c}
e^{i\ell\theta} r^{-\ell} e^{-br^2/4} \\[1ex]
0 \end{array}\right) \,,
\;\;\;\;\;\ \ell \leq 0
\label{eq38}
\end{equation}
and 
\begin{equation}
\Psi(r,\theta)  
=\left(\begin{array}{c}
e^{i\ell\theta} r^{-\ell +1} e^{-br^2/4} \\[1ex]
0 \end{array}\right) \,,
\;\;\;\;\;\ \ell \leq 0
\label{eq39}
\end{equation}
are zero modes of the Hamiltonian (\ref{eq36}). On the other hand, if $B= b r^{-3}$, with $b>0$, we have the following zero modes,
\begin{equation}
\Psi(r,\theta)  
=\left(\begin{array}{c}
r^{-\ell} e^{-b r^{-1}} e^{i\ell\theta}\\[1ex]
0 \end{array}\right) \,,
\;\;\;\;\;\ \ell \geq 1 \;,
\label{eq40}
\end{equation}
and 
\begin{equation}
\Psi(r,\theta)  
=\left(\begin{array}{c}
r^{-\ell+1} e^{-b r^{-1}} e^{i\ell\theta}\\[1ex]
0 \end{array}\right) \,,
\;\;\;\;\;\ \ell \geq 2 \;,
\label{eq41}
\end{equation}
Note, that both, (\ref{eq38}) and (\ref{eq40}), are zero modes of the single layer Hamiltonian (\ref{eq0}).
\\
Let us consider again the magnetic quantum dot defined in (\ref{eq33}). In particular, suppose that $b_1>0$ and $b_2>0$. 
Again, if an eigenstate of the Dirac equation exists in the two regions, the wave function should be continuous across the interface at $r = r_0$, that is, we should be able to match the eigenstates (\ref{eq38}) and (\ref{eq39}) with (\ref{eq40}) and (\ref{eq41}) respectively.
However, as we can see from these states, the spectrum of angular momentums in the in the region I,  does not have common values with the spectrum of angular momentums in the region II. Then, there are no eigenstates that coexist in both regions of the space. Let us consider now the possibility of superconfinement of the wave function,
i.e. the wave function different from zero only in one of the two regions.  
Let us concentrate on how to construct these states. 
We can suppose that the wave function is different from zero only in region I. Then, the wave function should be zero at $r=r_0$. We can construct a new zero mode of the Hamiltonian (\ref{eq36}) by linear combination of the eigenstates (\ref{eq38}) and (\ref{eq39}),
\begin{equation}
\Psi(r,\theta)  
=\left(\begin{array}{c}
e^{i\ell\theta} (r-r_0)r^{-\ell} e^{-br^2/4} \\[1ex]
0 \end{array}\right) \,,
\;\;\;\;\;\ \ell \leq 0
\label{eq42}
\end{equation}
For each value of $\ell$ satisfying $\ell \leq 0$, this state is a zero mode of the Hamiltonian (\ref{eq36}) and is vanished at $r=r_0$. Since, in view of (\ref{eq40}) and (\ref{eq41}),  there are no zero modes with $\ell \leq 0$ in the region II, we conclude that the wave function is confined in the region I. Outside of the region I the wave function is vanished. If we want to confine the wave function in the region II, we should combine linearly the eigenstates (\ref{eq40}) and (\ref{eq41}), so that the wave function is vanished at $r=r_0$ and $r=\infty$, 
\begin{equation}
\Psi(r,\theta)  
=\left(\begin{array}{c}
r^{-\ell} (r-r_0)e^{-b r^{-1}} e^{i\ell\theta}\\[1ex]
0 \end{array}\right) \,,
\;\;\;\;\;\ \ell \geq 2 \;,
\label{eq43}
\end{equation}
Thus, if the angular momentum is restricted to the condition $\ell \leq 0$, the wave function is superconfined into the region I, whereas if the angular momentum satisfies the condition $\ell \geq 2$, the wave function is superconfined into the region II. Notably, there are no zero modes with $\ell =1$.
\begin{figure}
\centering   
\includegraphics
[height=100mm]{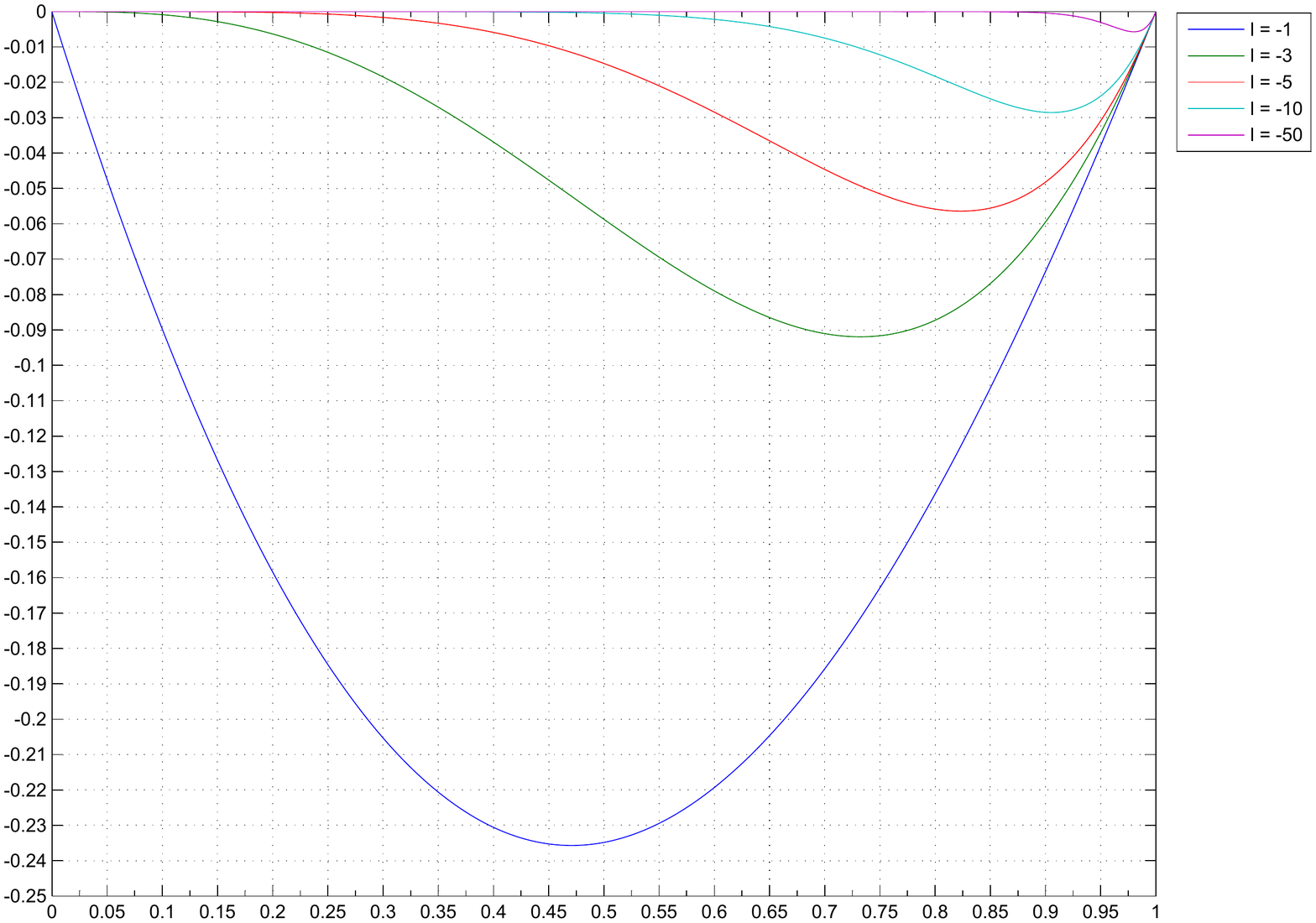} 
\caption{{
(Color online)`
{The zero mode field (\ref{eq42}) as a function of the radial coordinate $r$, for different values of $\ell$. From top to bottom, $\ell= -50,-10,-5,-3,-1 $.
}
}}
\label{tt2}
\end{figure}

\begin{figure}
\centering   
\includegraphics
[height=100mm]{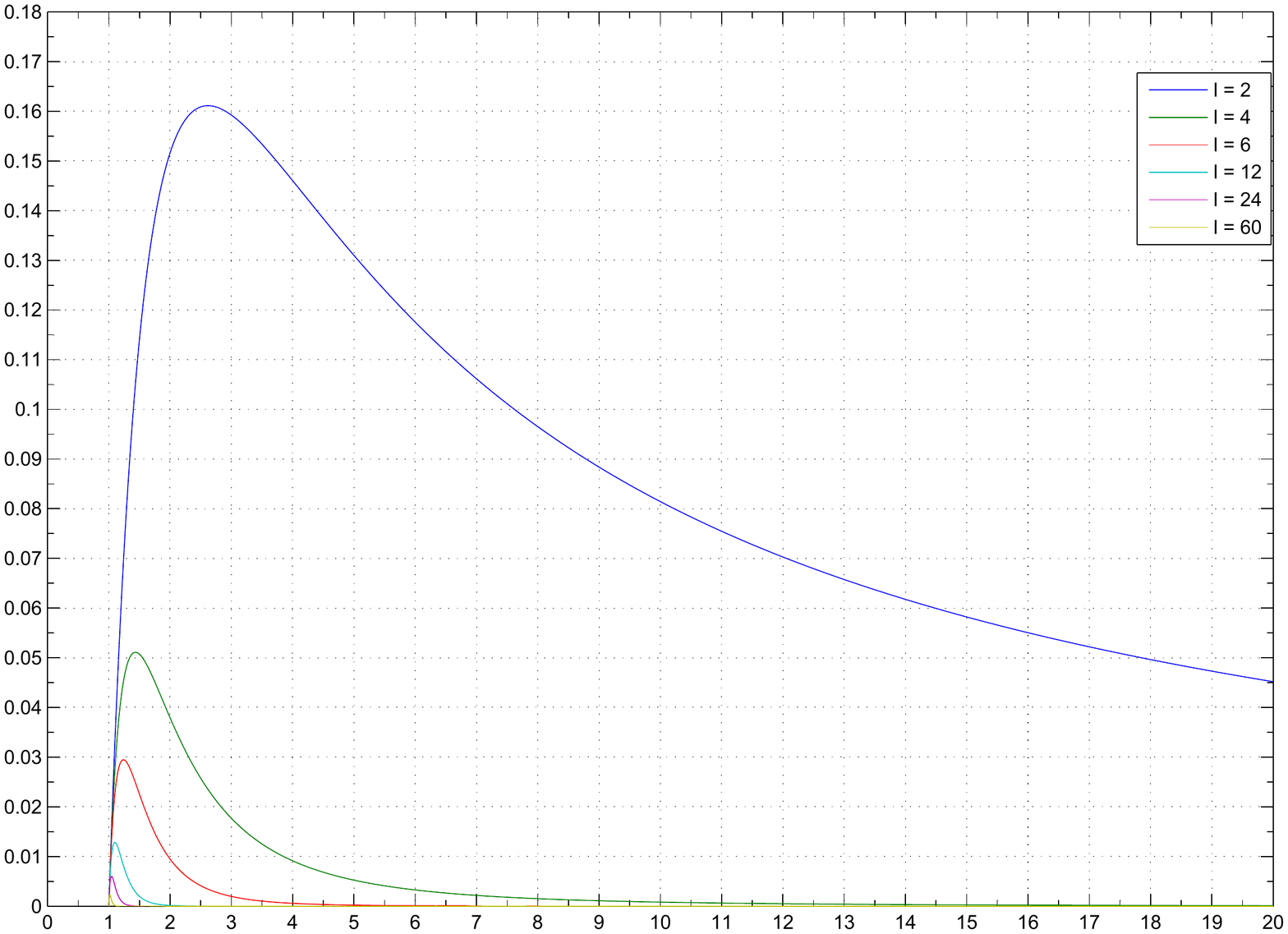} 
\caption{{
(Color online)
{The zero mode field (\ref{eq43}) as a function of the radial coordinate $r$, for different values of $\ell$. From top to bottom, $\ell= 2,4,6,12,24,60$.}
}}
\label{tt2}
\end{figure}
It is not difficult to generalize this result to a magnetic field configurations of the form, 
\begin{equation}
B(r) = \left\{\begin{array}{ll}
b_1r^{\alpha},\qquad &r<r_0\;\;\, \alpha>-2
\\[1.5ex]
b_2 r^{\alpha},\qquad &r>r_0\;\;\, \alpha<-2
\end{array}\right.\,.
\label{}
\end{equation}
\\
To conclude, we have described a new way of confining
Dirac-Weyl quasiparticles in graphene. By using a suitable magnetic field configurations, we have shown that the wave function may be restricted to a specific region of the space, being forbidden all transmission probability to the contiguous regions.
We hope that our
work can be useful in experimental realizations to the development of
mesoscopic structures based on graphene. 
From the theoretical point of view, we think that our work may be important to understand the behavior of Dirac fermions on magnetic field configurations. Finally, it would be interesting to extend our research to the excited energy levels. We expect to report on these issues in the future.

\vspace{0.6cm}

{\bf Acknowledgements}
\\
I would like to thank Charles Downing for helpful comments.
This work is supported by CONICET.

\end{document}